\documentclass[12pt]{iopart}

\usepackage{epsfig}% Include figure files
\usepackage{graphicx}% Include figure files
\usepackage{dcolumn}% Align table columns on decimal point
\usepackage{bm}% bold math

\begin{document}

\title{Effect of non-magnetic impurities on the magnetic states of anatase TiO$_2$}
\author{Sudhir K. Pandey \footnote{Electronic mail:
sk$_{_{-}}$iuc@rediffmail.com} and R. J. Choudhary}
%\altaffiliation{Electronic mail: sk$_{_{-}}$iuc@rediffmail.com}
%\author{R. J. Choudhary}
%\altaffiliation{Electronic mail: ram@csr.res.in}
%\affiliation{UGC-DAE Consortium for Scientific Research, University
%Campus, Khandwa Road, Indore - 452001, India}
\address{UGC-DAE Consortium for Scientific Research, University
Campus, Khandwa Road, Indore - 452001, India}

\date{\today}

\begin{abstract}
The electronic and magnetic properties of TiO$_2$, TiO$_{1.75}$,
TiO$_{1.75}$N$_{0.25}$, and TiO$_{1.75}$F$_{0.25}$ compounds have
been studied by using \emph{ab initio} electronic structure
calculations. TiO$_2$ is found to evolve from a wide-band-gap
semiconductor to a narrow-band-gap semiconductor to a half-metallic
state and finally to a metallic state with oxygen vacancy, N-doping
and F-doping, respectively. Present work clearly shows the robust
magnetic ground state for N- and F-doped TiO$_2$. The N-doping gives
rise to magnetic moment of $\sim$0.4 $\mu_B$ at N-site and $\sim$0.1
$\mu_B$ each at two neighboring O-sites, whereas F-doping creates a
magnetic moment of $\sim$0.3 $\mu_B$ at the nearest Ti atom. Here we
also discuss the possible cause of the observed magnetic states in
terms of the spatial electronic charge distribution of Ti, N and F
atoms responsible for bond formation.

\end{abstract}


\maketitle

\section{Introduction}
In recent years the scientific world is fascinated by the occurrence
of magnetism in non-magnetic materials where unpaired $d$ and/or $f$
electrons are absent. For example, materials like  Si\cite{erwin},
pyrolitic graphite, \cite{cervenka} fullerene, \cite{narymbetov}
CaO, \cite{elfimov, osorio} CaB$_6$,
\cite{monnier,maitiEPL,maitiPRL} SiC, \cite{zywietz} \emph{etc}.
contain only $s$ and $p$ elctrons and are reported to exhibit
magnetism. In the case of Si, magnetism is shown to arise from
surfaces \cite{erwin} whereas the observed room temperature
ferromagnetism (FM) in the pyrolitic graphite is attributed to the
two-dimensional networks of point defects.\cite{cervenka} The origin
of FM observed in CaB$_6$ has been a matter of controversy and two
schools of thought exist. One believes that it is arising from
magnetic impurities, whereas other has opinion that defects are
responsible for the FM.\cite{maitiPRL} Similarly, defects are also
found to be responsible for the creation of magnetic moments in CaO
and SiC compounds.\cite{osorio,volnianska} Thus the creation of
finite magnetic moments in these systems are attributed to the
uncompensated spins due to the surface effect, defects, and even to
magnetic impurities. In spite of these ambiguities, the phenomenon
has not only provided a new dimension to the spintronic based
materials but also throws a challenge to appreciate such an
occurrence from fundamental physics point of view as knowing the
exact cause for the creation of net magnetic moments and the nature
of interaction among them are non-trivial.

In this work we have attempted to explore the presence of
magnetism in the well studied non-magnetic semiconducting oxide
TiO$_2$ by creating oxygen vacancy and doping non-magnetic elements
at oxygen sites. In order to appreciate such an attempt, it is
important to note that getting ferromagnetic (FM) state in a
non-magnetic semiconductor by means of doping magnetic impurity has
attracted a great deal of attention in recent years and has emerged
as one of the important branch in the condensed matter physics and
materials science.\cite{ogaleAM,coey,ji,behan,ogalePRL,shinde}
However, the origin of ferromagnetism in such diluted magnetic oxide
semiconductors (DMOS) has been controversial, where the possibility
of the formation of secondary phase or clusters of magnetic impurity
cannot be ruled out. Recently a few reports have appeared which
emphasize induction of FM state in non-magnetic semiconducting
oxides by means of doping non-magnetic elements, creating defects or
changing oxygen
stoichiometry.\cite{pan,drera,kim,yangPRB,yang,venkatesan} Such DMOS
will indeed be quite advantageous, since then one does not need to
bother about issues related to incorporation of magnetic impurity in
the host matrix or its interaction with the carriers. Study of these
doped and undoped oxide semiconductors put forward a hope to
manipulate the opto-electronic devices by controlling their magnetic
properties. Thus our present attempt is expected to provide some
understanding towards the origin of magnetic moment in an otherwise
non-magnetic system and its possible applications in the spintronics
field.

TiO$_2$ is chosen for the study due to its well documented wide
range of optical, electrical and catalytic applications. TiO$_2$
based system also possesses good optical transmission in the visible
and near infra red regions making it a suitable candidate for the
magneto-optic device as well. For the present study we consider
three cases: (i) create oxygen vacancy (TiO$_{1.75}$), (ii) dope
nitrogen (TiO$_{1.75}$N$_{0.25}$), and (iii) dope fluorine
(TiO$_{1.75}$F$_{0.25}$) and examine their magnetic states. We take
these cases to compare the effect of non-magnetic anion doping of F
and N atoms with the aim to evaluate the influence of electron or
hole doping on the electronic and magnetic properties of the system.
Our results show that the oxygen vacancy does not induce any
appreciable magnetization in TiO$_2$. However, doping of N and F
atoms induces magnetic ground state, with magnetic moments appearing
prominently at N and Ti sites, respectively.

\section{Computational details}
The spin-unpolarized \emph{i.e.} non-magnetic (NM) and
spin-polarized (SP) electronic structure calculations of TiO$_2$,
TiO$_{1.75}$, TiO$_{1.75}$N$_{0.25}$, and TiO$_{1.75}$F$_{0.25}$
compounds have been carried out by using {\it state-of-the-art}
full-potential linearized augmented plane wave (FP-LAPW)
method.\cite{elk} The lattice parameters used in the calculations
are $a$ = 3.785 {\AA} and $c$ = 9.514 {\AA}; and the atomic position
taken for the oxygen atom is $z$ = 0.2066. The muffin-tin sphere
radii are chosen to be 2, 1.6, 1.5, and 1.45 a.u. for Ti, O, F, and
N atoms, respectively. For the exchange correlation functional, we
have adopted recently developed generalized gradient approximation
(GGA) form of Perdew {\em et al.}\cite{perdew} The self-consistency
was achieved by demanding the convergence of the total energy to be
smaller than 10$^{-4}$ Hartree/cell. The total magnitude of the
force per unit cell comes out to be less than 0.003 Hartree/Bohr.

\section{Results and discussions}
The anatase TiO$_2$ crystalizes in body center tetragonal structure
(space group \emph{I4$_1$/amd}) where Ti and O occupy 4$a$ (0,0,0)
and 8$e$ (0,0,$z$) Wyckoff positions, respectively. Therefore, the
conventional unit cell contains 4 Ti and 8 O atoms. The atomic
arrangement in the unit cell is shown in Fig. 1. Each Ti atom is
surrounded by 6 O atoms. Out of 6 Ti-O bonds, 4 bonds are of equal
length (1.937 {\AA}) and rest two have bondlength of 1.965 {\AA}. In
order to see the effect of non-magnetic electron and hole doping on
the electronic and magnetic properties of the compound we have
replaced O2 by fluorine and nitrogen atoms, respectively. Similarly,
the effect of oxygen vacancy is also studied by removing O2 atom.

The total density of states (TDOS) per formula unit (fu) of TiO$_2$,
TiO$_{1.75}$, TiO$_{1.75}$N$_{0.25}$, and TiO$_{1.75}$F$_{0.25}$
compounds are shown in Fig. 2. The TDOS of TiO$_2$ in both the spin
channels are perfectly symmetric which is in accordance with the
non-magnetic ground state of the compound. The band gap of TiO$_2$
is found to be $\sim$2.1 eV, which is about 1 eV lower than
experimentally observed value.\cite{ogaleAM} The discrepancy between
the calculated and experimental results is not surprising as GGA
calculations are often found to underestimate the band gap of
semiconductors and insulators. The creation of oxygen vacancy does
not have much influence on the magnetic state of the system as TDOS
of TiO$_{1.75}$ in both the spin channels also look symmetric as
evident from Fig. 2(b). This observation is in contrast to the
previous pseudo-potential based calculations showing magnetic ground
state driven by oxygen vacancy.\cite{yangPRB} The GGA+$U$ results of
Yang \emph{et al}. show the formation of magnetic moment of 1
$\mu_B$ each at two neighbouring Ti sites.\cite{yangPRB} In order to
look into the possibility that on-site Coulomb interaction ($U$)
among Ti 3$d$ electrons might be responsible for the creation of
magnetic moments in the work of Yang \emph{et al}, we have also
carried out GGA+$U$ calculation and could not find any magnetic
moment at the Ti sites even for $U$=4 eV which is a reasonable value
of $U$ for the 3$d$ electrons.\cite{sudhirLCO,sudhirSRNO,sudhirSCO}
At this juncture, it is important to note that full-potential
calculations are often found to provide more accurate results in
comparison to pseudo-potential ones. Our calculations give
negligibly small magnetic moment at Ti sites, which is consistent
with earlier full-potential based result.\cite{kim} Interestingly,
oxygen vacancy creates an electron-like band just below the bottom
of the conduction band (CB). The band gap between top of the valance
band (VB) and bottom of the impurity band is found to be $\sim$1.1
eV and that between top of the impurity band and bottom of the
conduction band is $\sim$0.1 eV. This result suggests that the
oxygen vacancy gives rise to significantly large increment in the
number of charge carriers at room temperature, which is consistent
with the available experimental data.\cite{ogaleAM}

The doping of non-magnetic impurities at O2 site provides highly
asymmetric energy distribution of the states closer to the Fermi
level ($\epsilon_F$) in both the spin channels as evident from Figs.
1(c) and 1(d). This suggests the creation of net magnetic moments in
the compounds. The ground state of N-doped TiO$_2$ is found to be
magnetic as the energy of SP solution is $\sim$102 meV/fu less than
that of NM solution. The N-doping shifts the edge of the valence
band towards $\epsilon_F$ by $\sim$0.5 eV. Moreover, it gives rise
to extended VB states. As expected N-doping provides hole-like
impurity band. Interestingly, the ground state of N-doped TiO$_2$ is
found to be half-metallic as there is no density of states (DOS) at
the $\epsilon_F$ in the up-spin channel and the band gap in this
channel is found to be $\sim$2 eV. The F-doping also gives rise to
magnetic ground state as the energy of SP solution is $\sim$71
meV/fu less than that of NM solution. The VB of F-doped TiO$_2$ is
extended over a wide energy range with the appearance of new states
between -10.9 and -8.8 eV which were absent in earlier cases. These
states are expected to arise from the fluorine atoms, which becomes
clear in later part of the manuscript. As opposed to the case of
oxygen vacancy, the electron-like impurity band of F-doped TiO$_2$
is found to overlap with the CB minimum resulting in metallic ground
state as evident from Fig. 1(d). Thus our results provide the simple
but elegant way of controlling the electronic transport of TiO$_2$
as it evolves from a wide-band-gap semiconductor to a
narrow-band-gap semiconductor to a half-metallic state and finally
to a metallic state with oxygen vacancy, N-doping and F-doping,
respectively.

In order to know the energy distribution of different states we have
plotted the Ti 3$d$ and O 2$p$ partial density of states (PDOS) of
TiO$_2$ and TiO$_{1.75}$ in Fig. 3. It is evident from the Fig. 3(a)
that the states closer to the top of the VB have dominating
contribution from O 2$p$ orbitals, whereas states in the vicinity of
the bottom of VB has mixed Ti 3$d$ and O 2$p$ character. The bottom
of the CB consists of Ti 3$d$ states. Moreover, one can see the
finite O 2$p$ PDOS in the CB region above 2 eV. The finite
contribution of Ti 3$d$ and O 2$p$ orbitals in the VB and CB,
respectively suggests the breakdown of the pure ionic model for Ti-O
bonds and can be considered as a signature of strong hybridization
between Ti 3$d$ and O 2$p$ states. The effect of oxygen vacancy on
the electronic states of TiO$_2$ is displayed in Fig. 3(b) where we
have plotted Ti 3$d$ PDOS of Ti1 and Ti2 atoms which have got most
influenced by removal of O2 atom. Two obvious features are visible:
(i) the bottom of the VB corresponds to Ti2 3$d$ states which shifts
by $\sim$0.12 eV deeper in comparison to Ti1 3$d$ states and (ii)
the contribution of Ti2 3$d$ states to the impurity band is about
75\%. The first one may be attributed to the more hybridization of
Ti2 3$d$ and O2 2$p$ orbitals in comparison to Ti1 3$d$ and O2 2$p$
orbitals, as the bondlength of Ti2-O2 is $\sim$0.03 {\AA} lesser
than that of Ti2-O1, resulting in increased separation between
bonding and antiboding molecular orbitals. The second one can be
attributed to the orbital polarization. In octahedral symmetry two
$d$ electrons of the elemental Ti is expected to occupy $t_{2g}$
orbitals. In the anatase structure shown in Fig. 1, only $d_{xy}$
orbital is expected to show significant overlap with O2 2$p$
orbital, resulting in transfer of electron to O2 and making it
negatively charged. In the case of vacancy at O2 site, this
transferred electron would be taken back by Ti2 and would mainly
contribute to the impurity band as seen from the calculation.

The N-doping at O2 site does not have any influence on the magnetic
state of Ti atoms. The 2$p$ PDOS of N, O1 and O3 are shown in Fig.
4(a) as these states are found to be most influenced by the doping.
The distribution of these states is highly asymmetric in both the
spin channels suggesting of net magnetic moments at N, O1 and O3
atoms. The magnetic moment at N site is found to be $\sim$0.4
$\mu_B$ and that at O1 and O3 sites are $\sim$0.1 $\mu_B$ each. The
contribution of magnetic moments from rest of the sites is
negligibly small. The total magnetic moment per formula unit comes
out to be $\sim$0.25 $\mu_B$. The distribution of impurity PDOS
around the $\epsilon_F$ provides valuable information about the
nature of magnetic interaction among the magnetic moments of the
impurity atoms.\cite{sato} In the up-spin channel one can see the
absence of states closer to the $\epsilon_F$. However, in the
down-spin channel there are finite states in the vicinity of
$\epsilon_F$. These states are of highly mixed N 2$p$ and O 2$p$
characters. One can also see the appearance of a sharp peak just
above the $\epsilon_F$ which has dominating contribution from N 2$p$
orbitals.

Contrary to the N-doping, the F-doping  at O2 site gives rise to net
magnetic moment of $\sim$0.3 $\mu_B$ for Ti2 atom keeping magnetic
moments of rest of the atoms negligibly small. The total magnetic
moment per formula unit comes out to be $\sim$0.11 $\mu_B$/fu. In
Fig. 4(b) we have shown the Ti2 3$d$, O1 2$p$ and F 2$p$ PDOS as
these states are found to be most influenced by the F-doping.
Interestingly, the electron type impurity band has mainly Ti2 3$d$
character. The energy distribution of this band in both the spin
channels is highly asymmetric and responsible for creation of
magnetic moment at Ti2 site. The states found below -8.8 eV are
mainly arising from F 2$p$ orbitals and O1 2$p$ orbitals are mainly
contributing between -7.8 to -2.9 eV.

The above results clearly establish the magnetic ground state on
doping non-magnetic impurities at the O sites. However, N-doping
creates magnetic moments at N site whereas F-doping gives rise to
magnetic moment at Ti2 site. Entirely different effect of N- and F-
doping at the magnetic moments is unusual and appears to be related
with the spatial electronic charge distribution of Ti2, N and F
atoms responsible for bond formation. Pure TiO$_2$ is non-magnetic
as the valance electrons are equally populated in both the spin
channels. The F 2$p$ orbitals are highly localized as evident from
Fig. 4(b). Such localization of $p$ orbitals prohibits sharing of
electrons from the neighboring Ti2 3$d$ electrons and making
transfer of 3$d$ electrons to F site energetically more favorable.
The F 2$p$ orbitals are more than half-filled and hence only
down-spin electrons are allowed to be transferred resulting in net
spin polarization of Ti2 3$d$ electrons. Contrary to the F 2$p$
orbitals, N 2$p$ orbitals are extended which can easily share its
electrons with neighboring Ti2 3$d$ orbitals making Ti2-N bonds more
covalent in nature. In covalent bond there are equal probability of
sharing the up- and down-spin electrons. Moreover, N 2$p$ orbitals
contains one more electron in comparison to Ti2 3$d$ electrons. Thus
the sharing of Ti2 3$d$ and N 2$p$ electrons is expected to provide
net spin polarization at N site due to inert 2$p$ electron.

Finally, we explore the possibility of utilization of the above
results in the real systems. First hindrance appears to be the
synthesis of such compounds as the dopant percentage (either N or F)
studied here is quite high if we compare with the results of
ZnO.\cite{osorio} However, Chen \emph{et al.} have reported to
synthesize as high as 15\% N-doped TiO$_2$.\cite{chen} Since the
difference between ionic radius of O and N ions is almost the same
as that between O and F ions, therefore, F-doped compound is also
expected to be synthesized easily. Moreover, we performed
calculations on 6.125\% N- and F- doped TiO$_2$ and found similar
results as far as formation of magnetic moments are concerned. Thus
it appears that the synthesis of compounds which shows finite
magnetic moments is not a problem. However, the mere formation of
finite magnetic moments at different sites does not guarantee the
collective magnetism which is essential for any practical
application. To ascertain this one needs to estimate the optimal
concentration of dopant required for percolation of the magnetic
interaction. Any such attempt with fair accuracy is very much time
consuming under FP-LAPW method and beyond the scope of the present
work.  However, we feel that the dopant percentage studied here is
high enough to induced collective magnetism and motivate other
workers to look into this aspect.

\section{Conclusions}
In conclusion, we have examined the effect of oxygen vacancy and
doping of non-magnetic anion in TiO$_2$ on its magnetic properties.
It turns out that oxygen vacancy leads to electron doping in the
system, though it does not induce appreciable magnetic moment in the
system. Interestingly, doping of N gives rise to robust
ferromagnetic half-metallic ground state for TiO$_{1.75}$N$_{0.25}$
with the appearance of magnetic moment at N and two neighboring O
sites and hence expected to provide an innovative prospect for a
novel class of DMOS material with anion doping. On the other hand
the ground state of F-doped sample is found to be metallic with the
induction of magnetic moment at Ti site nearest to F atom. Since we
observe that the electronic state of TiO$_2$ can be controlled from
wide-band-gap semiconductor to narrow-band-gap semiconductor to half
metallic to metallic states by selectively managing the anion site,
it gives us a hope to tune the band gap of TiO$_2$ to further extend
its application domain in photo-catalytic and magneto-optic devices.
Such a study would also provide us an understanding of the origin of
magnetism in a semiconducting oxide system by non-magnetic anion
doping.

%\section{Acknowledgements}

\begin{figure}
  \includegraphics[width=8cm]{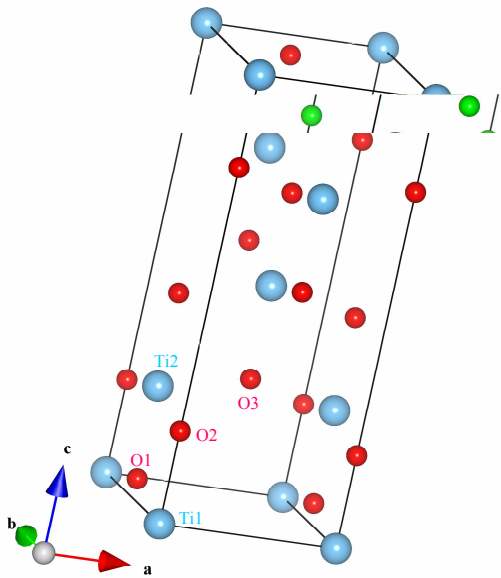}\\
  \caption{(Color online) Atomic arrangements of TiO$_2$ in the unit
cell.}\label{1}
\end{figure}

\begin{figure}
  \includegraphics[width=8cm]{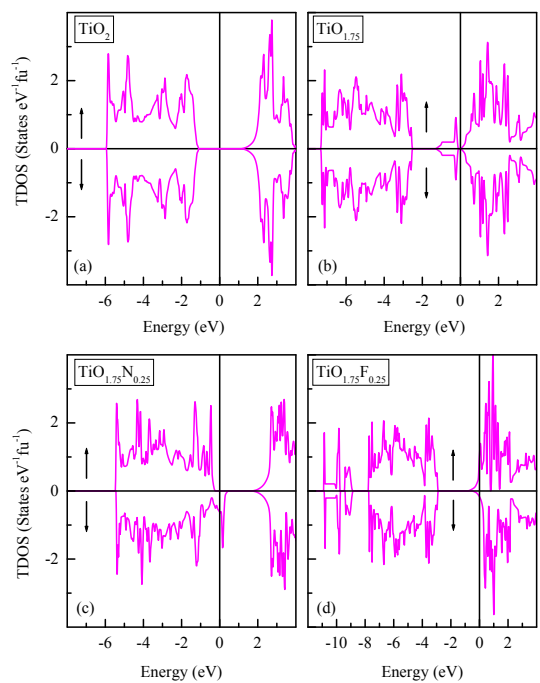}\\
  \caption{(Color online) Total density of states (TDOS) per formula
unit of (a) TiO$_2$, (b) TiO$_{1.75}$, (c) TiO$_{1.75}$N$_{0.25}$,
and (d) TiO$_{1.75}$F$_{0.25}$ compounds.}\label{2}
\end{figure}

\begin{figure}
  \includegraphics[width=8cm]{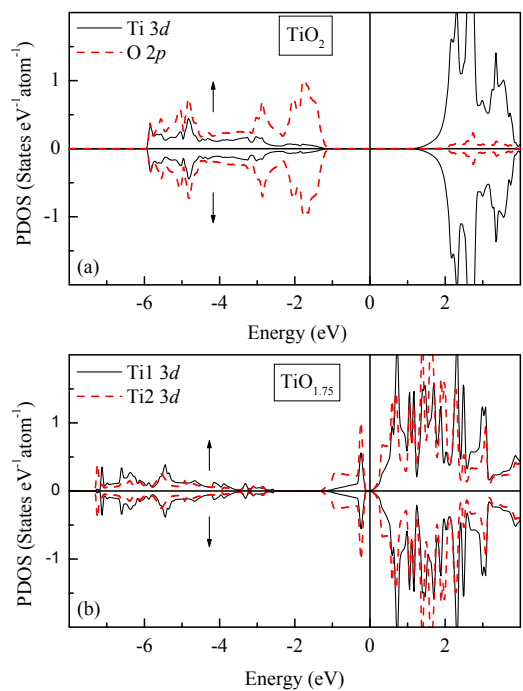}\\
  \caption{(Color online) Partial density of states (PDOS) of (a) Ti
3$d$ and O 2$p$ states in TiO$_2$ and (b) Ti1 3$d$ and Ti2 3$d$
states in TiO$_{1.75}$.}\label{3}
\end{figure}

\begin{figure}
  \includegraphics[width=8cm]{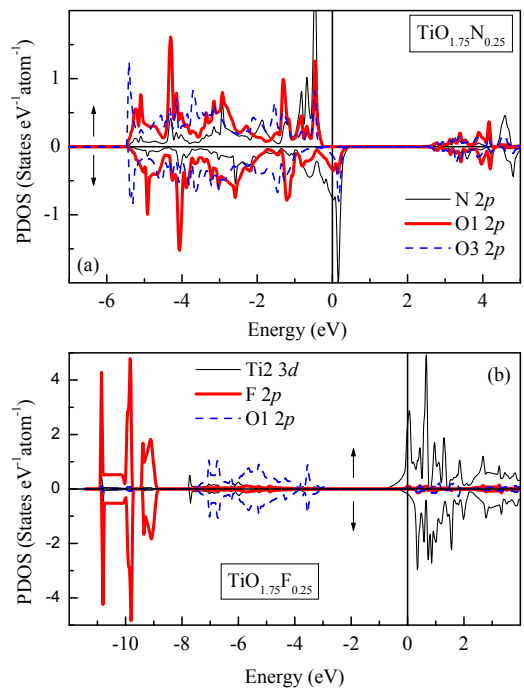}\\
  \caption{(Color online) Partial density of states (PDOS) of (a) N
2$p$, O1 2$p$ and O3 2$p$ states in TiO$_{1.75}$N$_{0.25}$ and (b)
Ti2 3$d$, F 2$p$ and O1 2$p$ states in
TiO$_{1.75}$F$_{0.25}$.}\label{3}
\end{figure}

%\end{document}

%\section{Figure Captions:}

\pagebreak

%\section{Tables}

%Table 1:

\vspace{2ex}
%\begin{ruledtabular}

%\end{ruledtabular}

\end{document}